\documentclass[11pt]{article}

\usepackage[margin=1.1in]{geometry}
\usepackage{graphicx}
\usepackage{tikz}
\usetikzlibrary{arrows.meta}

\begin{document}

\title{Extrapolating Solution Paths of Polynomial Homotopies \\
       towards Singularities with PHCpack and phcpy\thanks{Supported by the 
National Science Foundation under grant DMS 1854513.}}

\author{Jan Verschelde\thanks{University of Illinois at Chicago,
Department of Mathematics, Statistics, and Computer Science,
851 S. Morgan St. (m/c 249), Chicago, IL 60607-7045.
Email: {\tt janv@uic.edu}, URL: {\tt http://www.math.uic.edu/$\sim$jan}.}
\and Kylash Viswanathan\thanks{University of Illinois at Chicago,
Department of Mathematics, Statistics, and Computer Science,
851 S. Morgan St. (m/c 249), Chicago, IL 60607-7045.
Email: {\tt kviswa5@uic.edu}.}}

\date{1 May 2024}

\maketitle

\begin{abstract}
PHCpack is a software package for polynomial homotopy continuation,
which provides a robust path tracker [Telen, Van Barel, Verschelde, SISC 2020].
This tracker computes the radius of convergence of Newton's method,
estimates the distance to the nearest path, and
then applies Pad\'{e} approximants to predict the next point on the path.
A priori step size control is less sensitive to finely tuned tolerances 
than a posteriori step size control, and is therefore robust.
The Python interface phcpy is extended with a new step-by-step tracker
and is applied to experiment with extrapolation methods
to accurately locate the singular points at the end of solution paths.

\end{abstract}
\section{Introduction}

A polynomial homotopy is a family of polynomial systems which depend
on one parameter~$t$.  Regular solutions of a polynomial homotopy have
Taylor series developments in~$t$.  The solutions paths are analytic
functions of~$t$.  Application of the theorem of Fabry~\cite{Fab1896}
enables the location of the nearest singular solution.
The calculations in this paper can be considered in the area of
numerical analytic continuation~\cite{Hen66,Tre20,Tre23}.

Extrapolation methods~\cite{BZ91,Sid03,Wen89},
in particular Aitken's algorithm and the rho algorithm~\cite{Wyn56},
are effective in accelerating logarithmically converging series
towards a single pole of a polynomial homotopy.
We demonstrate the interplay between compiled code in a library,
for the computationally intensive calculation of the Taylor series, and 
the interactive Python scripts for the extrapolation methods.
Our calculations happen in Jupyter notebooks~\cite{Jupyter}
with phcpy~\cite{OFV19,Ver14} in a Python kernel.
The interface to the compiled code in PHCpack~\cite{Ver99}
does not require compilation as it is done through the Ctypes module 
of Python which allows to call functions in libraries directly,
similar to ccall in Julia.

Since version 2.4.88, PHCpack became a crate of alire\footnote{Version
1.2.2 of alr, GNAT 12.2.1 and gprbuild 22.0.1.}
the package manager of the gnu-ada compiler, 
which builds the executable phc and libPHCpack, integrating code
of MixedVol~\cite{GLW05} and DEMiCs~\cite{MT08}, 
for fast mixed volume computation, and algorithms for multiple 
double arithmetic of QDlib~\cite{HLB01} and CAMPARY~\cite{JMPT16}.
The development of phcpy was motivated in part by SageMath~\cite{ES10}.
All software is free and open source,
released under version~3 of the GNU GPL license.
Computations are done with phcpy~1.1.4
and version 2.4.90 of PHCpack.

The main contribution of this paper is the illustration of phcpy
to visualize solution paths and to experiment with extrapolation methods.

\section{Extrapolation Experiments}

\begin{figure}[hbt]
\centerline{\includegraphics[width=13cm]{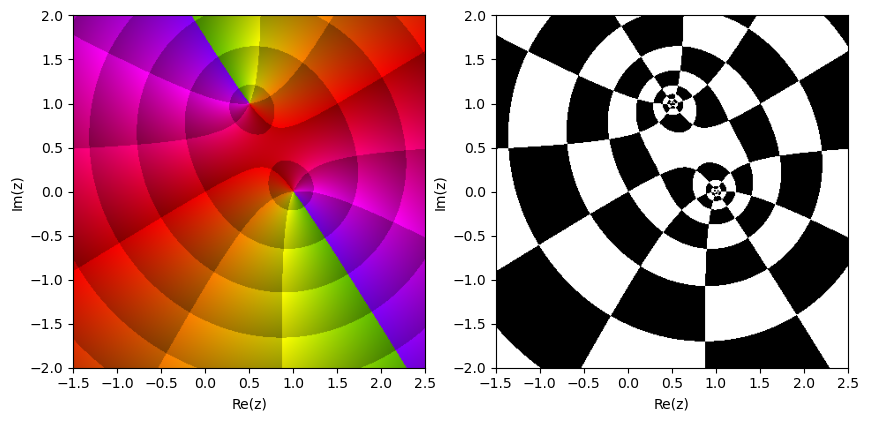}}
\caption{A phase portrait of 
         $f(z) = \sqrt{4/5(1/2-I)(1-z)(1/2+I-z)}$, depicting at the left the
color-coded phase $f/|f|$ on the domain of~$f$.
At the right is a polar chessboard.}
\label{figphaseportrait}
\end{figure}

The polynomial homotopy
\begin{equation} \label{eqfirsthomotopy}
  x^2 - \frac{4}{5} \left( \frac{1}{2} - I \right)
        \left( \vphantom{\frac{1}{2}} 1 - t \right)
        \left( \frac{1}{2} + I - t \right) = 0, \quad I = \sqrt{-1}.
\end{equation}
has two singularities: one at $t = 1$ and another at $t = 1/2 + I$.
The coefficient $4/5(1/2 - I)$ makes that at $t=0$, the solution
paths $x(t)$ start at $x(0) = \pm~1$.
The phase portrait~\cite{Weg12},
made with {\tt complexplorer-0.1.2}~\cite{Kuv23}
(which depends on {\tt matplotlib}~\cite{Hun07}),
is shown in Figure~\ref{figphaseportrait}.

Using the algorithms in~\cite{BV18}
to compute the Taylor series is done with phcpy in the code snippet below:
\begin{verbatim}
from phcpy.solutions import make_solution
from phcpy.series import double_newton_at_point

pol = ['x^2 - (4/5)*(1/2 - I)*(1 - t)*(1/2 + I - t);']
variables = ['x', 't']
sols = [make_solution(variables, [1, 0])]
deg = 33  # degree of truncation
nit = 8   # number of iterations of Newton's method
srs = double_newton_at_point(pol, sols, idx=2, maxdeg=deg, nbr=nit)
\end{verbatim}
The {\tt srs} contains the string representation of the Taylor series
and the coefficients can be extracted with the help of SymPy~\cite{SymPy11}.
Denoting $c_n$ as the coefficient of $t^n$ in the Taylor series,
the ratio $c_{n}/c_{n+1}$
equals {\tt (1.0362677867627397} {\tt -0.03656143770249911j)},
for $n=31$, illustrating the very slow convergence to~1.
Sequences such as this are said to be converge logarithmically,
as it may take about twice as many terms in the series to gain
one bit of accuracy.

The rho algorithm~\cite{Wyn56} performs spectacularly well on 
the Taylor series of $x(t) = \sqrt{1-t}$, where there is only one pole.
In the definition of the function {\tt rhoComplex} below, 
observe the inverse divided differences.
The connection with Thiele interpolation~\cite{CW87} is one possible
justification of its good performance on $x(t) = \sqrt{1-t}$.
The case of nearby poles is analyzed in~\cite{VV24b}.

\begin{verbatim}
def rhoComplex(nbr):
    """
    Runs the rho algorithm in complex double arithmetic,
    on the numbers given in the list nbr, using x(n) = n+1.
    Returns the last element of the table of extrapolated numbers.
    """
    rho1 = [1.0/(nbr[n] - nbr[n-1]) for n in range(1, len(nbr))]
    rho = [nbr, rho1]
    for k in range(2, len(nbr)):
        nextrho = []
        for n in range(k, len(nbr)):
            invrho1 = complex(k)/(rho[k-1][n-k+1] - rho[k-1][n-k])
            nextrho.append(rho[k-2][n-k+1] + invrho1)
        rho.append(nextrho)
    return nextrho[-1]
\end{verbatim}

Because the two singularities of~(\ref{eqfirsthomotopy})
are relatively too close to each other, the rho algorithm
fails to improve the sequence of Fabry ratios.
In the second homotopy:
\begin{equation} \label{eqsecondhomotopy}
    x^2 - \left( \frac{1}{272} \right)
      \left( \vphantom{\frac{1}{2}} -4 - 16 I \right)
      \left( \vphantom{\frac{1}{2}} 1 - t \right)
      \left( \vphantom{\frac{1}{2}} -4 + 16 I - t \right) = 0,
\end{equation}
the singularity at $-4 + 16 I$ is much farther 
from the other singularity at~$1$, and then the rho algorithm ends at
{\tt (1.0000000000014202+4.354118985724171e-14j)}, using 32 coefficients
of the Taylor series, with an error of {\tt 1.42e-12}.

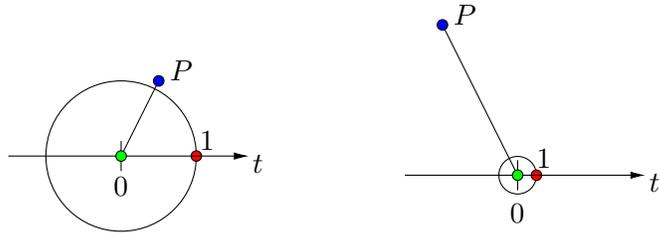
\begin{figure}[hbt]
\begin{center}
\begin{picture}(270,90)(0,0)
\put(0,0){
\begin{tikzpicture}
\draw (0cm, 0cm) circle(1cm);
\draw (0mm, -2mm) -- (0mm, 2mm);
\draw (0mm, 0mm) -- (5mm, 10mm);
\draw[fill=red] (1cm, 0cm) circle(0.7mm);
\draw[fill=blue] (5mm, 10mm) circle(0.7mm);
\draw[-{Latex[length=2mm, width=1mm]}] (-15mm, 0mm) -- (17mm, 0mm);
\node[text width=5mm] at (9mm, 11.34mm) {$P$};
\node[text width=5mm] at (20mm, -1mm) {$t$};
\node[text width=5mm] at (13mm, 2mm) {$1$};
\node[text width=5mm] at (1.5mm, -4mm) {$0$};
\draw[fill=green] (0mm, 0mm) circle(0.7mm);
\end{tikzpicture}
}
\put(150,0){
\begin{tikzpicture}
\draw (0cm, 0cm) circle(0.25cm);
\draw (0mm, -2mm) -- (0mm, 2mm);
\draw (0mm, 0mm) -- (-10mm, 20mm);
\draw[fill=red] (0.25cm, 0cm) circle(0.7mm);
\draw[fill=blue] (-10mm, 20mm) circle(0.7mm);
\draw[-{Latex[length=2mm, width=1mm]}] (-15mm, 0mm) -- (17mm, 0mm);
\node[text width=5mm] at (-6mm, 21.34mm) {$P$};
\node[text width=5mm] at (20mm, -1mm) {$t$};
\node[text width=5mm] at (5mm, 2.2mm) {$1$};
\node[text width=5mm] at (1.5mm, -5mm) {$0$};
\draw[fill=green] (0mm, 0mm) circle(0.7mm);
\end{tikzpicture}
}
\end{picture}
\end{center}
\caption{At the left, the pole $P = 1/2 + I$ is relatively close to~1,
while at the right, the pole $P = -4 + 16 I$ is much farther from~1.}
\label{figtwopoles}
\end{figure}

The differences between the location of the singularities
in the homotopies~(\ref{eqfirsthomotopy}) and~(\ref{eqsecondhomotopy})
is shown in the schematic of Figure~\ref{figtwopoles}.
The $P$ at the right of Figure~\ref{figtwopoles} illustrates
the notion of {\em the last pole}, introduced in~\cite{VV22}.

The right of Figure~\ref{figtwopoles} is a representative schematic
for a solution path converging to a singular solution at~1.
The effect a nearby and far away poles on extrapolation method
is the subject of~\cite{VV24b}.

\section{Path Tracking Towards a Singular Solution}

One important innovation in the modernization~\cite{Ver14} of PHCpack
with a scripting interface was the introduction of a step-by-step
path tracker, which allows the user to ask for the next point on a path.
Recently, another step-by-step tracker 
which applies the a priori step size control algorithms of~\cite{TVV20}
was added to phcpy.

\begin{figure}[hbt]
\centerline{\includegraphics[width=13cm]{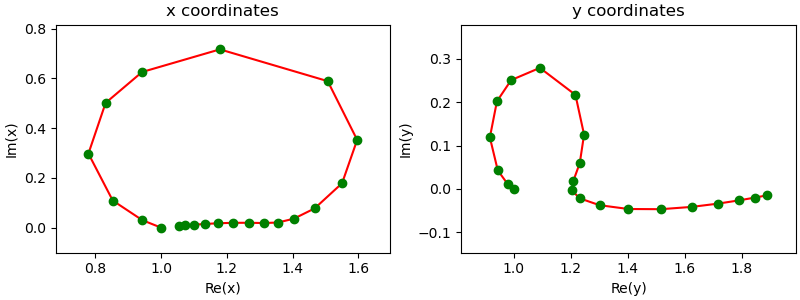}}
\caption{Points on one solution path starting at~$(1, 1)$
and converging to~$(1, 2)$.}
\label{figojika1path}
\end{figure}

The homotopy
\begin{equation} \label{eqojika1homotopy}
   \gamma (1 - t)
   \left(
     \begin{array}{rcl}
        x^2 - 1 & = & 0 \\
        y^2 - 1 & = & 0 \\
     \end{array}
   \right)
   + t
   \left(
     \begin{array}{rcl}
        x^2 + y - 3 & = & 0 \\
        x + 0.125 y^2 - 1.5 & = & 0
     \end{array}
   \right)
\end{equation}
ends at $t=1$ at an example of~\cite{Oji87a},
which has a triple root at~$(x, y) = (1, 2)$.
Figure~\ref{figojika1path} shows one path defined by
the homotopy~(\ref{eqojika1homotopy}), converging to~$(1, 2)$.

The code below collects the points on the solution path
in the list {\tt path}.
The constant~$\gamma$ in the homotopy~(\ref{eqojika1homotopy}) is
set to the value $-0.917 - 0.398 I$.
Expecting a singularity at the end of the path,
as long as the distance between the closest pole and~1.0
is larger than {\tt 1.0e-4}, the path tracker continues.
Three lists are produced: in {\tt path} are the points on the path,
in {\tt predicted} are the predicted solutions, and {\tt poles}
contains the list of poles closest to the path.

\begin{verbatim}
from phcpy.solutions import make_solution
from phcpy.curves import set_gamma_constant
from phcpy.curves import initialize_double_artificial_homotopy
from phcpy.curves import set_double_solution, get_double_solution
from phcpy.curves import double_predict_correct, double_t_value
from phcpy.curves import double_closest_pole

pols = ['x**2+y-3;', 'x+0.125*y**2-1.5;']
start = ['x**2 - 1;', 'y**2 - 1;']
startsol = make_solution(['x', 'y'], [1, 1])
set_gamma_constant(complex(-0.917, -0.398))
initialize_double_artificial_homotopy(pols, start)
set_double_solution(2, startsol)
path = [startsol]
predicted = []
poles = []
cfp = complex(0.0)
while abs(cfp - 1.0) > 1.0e-4:
    double_predict_correct()
    (repole, impole) = double_closest_pole()
    tval = double_t_value()
    cfp = complex(tval + repole, impole)
    poles.append(cfp)
    sol = get_double_solution()
    predsol = get_double_predicted_solution()
    path.append(sol)
    predicted.append(predsol)
print(sol)
\end{verbatim}

\noindent The code prints
\begin{verbatim}
    t :  9.99968379127468E-01   0.00000000000000E+00
    m : 1
    the solution for t :
     x :  1.05353846638456E+00   6.88135807075172E-03
     y :  1.89010705713382E+00  -1.44977473137858E-02
    == err :  6.369E-14 = rco :  7.184E-04 = res :  2.849E-17 =
\end{verbatim}
The value {\tt 7.184E-04} next to {\tt rco} is the estimate
for the inverse condition number of the Jacobian matrix at the point,
which gives an upper bound on the error of the update in Newton's method.
In particular, the update of Newton's method is correct up to the
last four decimal places.  Even as {\tt 0.999684} is already close
to~1, the solution is thus still well conditioned.

The number {\tt (1.0000808949264557-6.886127445687259e-08j)}
ends the list {\tt poles}.
Observe how small the imaginary part is.
Only seven terms in the Taylor series
were used to compute this approximation for~1.0.

\begin{figure}[t!]
\centerline{\includegraphics[width=13cm]{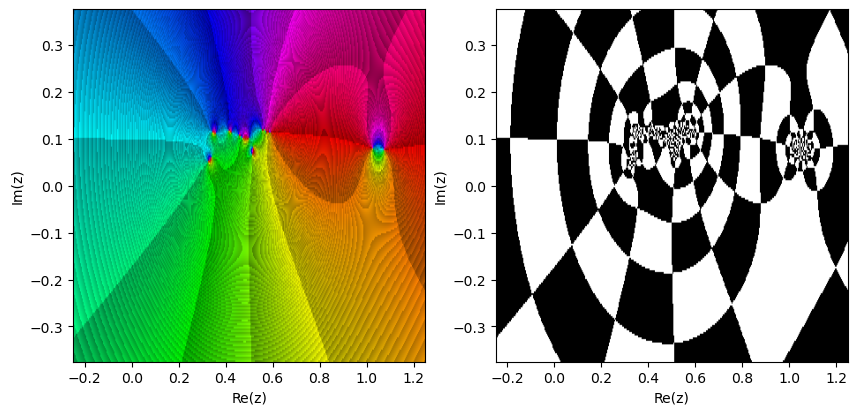}}
\caption{A phase portrait of $f(z)$, defined in~(\ref{eqojika1poles})
by the poles of one path converging to a singular solution.}
\label{figphaseportraitojika1path}
\end{figure}

To visualize the poles with {\tt complexplorer},
consider the function
\begin{equation} \label{eqojika1poles}
   f(z) = \sum_{p \in L} \frac{1}{z - p},
\end{equation}
where $L$ is the list of the first twelve poles.
The phase portrait of~$f$
is shown in Figure~\ref{figphaseportraitojika1path}.
The phase portrait starts to look interesting 
at the first closest pole, around $t \approx 0.471 + 0.101 I$.

\begin{figure}[ht!]
\centerline{\includegraphics[width=13cm]{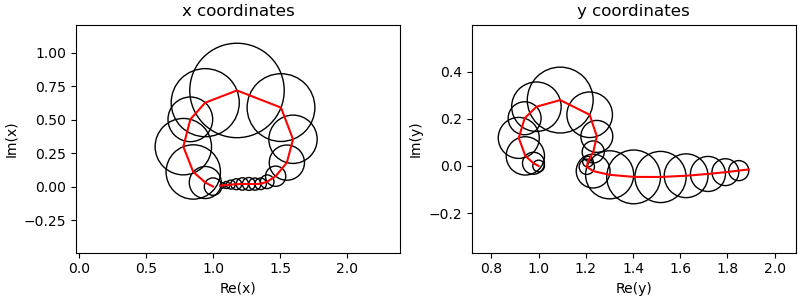}}
\caption{At each point on the path, a circle is drawn,
with center at the point and the radius is the distance to
the predicted solution.}
\label{figojika1circles}
\end{figure}

Figure~\ref{figojika1circles} is constructed using the information
of the points on the path and the predicted solutions.
The predicted solutions are evaluated Pad\'{e} approximants,
evaluated in the radius of convergence for Newton's method,
as estimated by the application of the theorem of Fabry.
The plots in Figure~\ref{figojika1circles} show a sequence of
circles with overlapping interiors.  Towards the end of the path,
observe that the radii of the circles decrease, as we approach
the singularity at the end.  The smaller circles in the middle
of the path indicate nearby poles.

\section{Iterated Aitken Extrapolation}

Looking at the list {\tt path},
the solution at the end is not very accurate:
the 1-norm of its solution components equals {\tt 2.27e-01},
which means that we have only one decimal place of accuracy.
If we consider the last seven points on the path,
then we observe a very slowly converging sequence of points.
Aitken extrapolation is effective in accelerating logarithmically
converging sequences~\cite{Kow81}.
The function below contains the definition of Aitken extrapolation:

\begin{verbatim}
     def Aitken(x):
         """
         Applies Aitken extrapolation to the sequence x.
         Returns the transformed sequence.
         """
         y = [0.0 for _ in range(len(x)-2)]
         for k in range(len(x)-2):
             dxk = x[k+1] - x[k]
             ddx = x[k+2] - 2*x[k+1] + x[k]
             y[k] = x[k] - dxk*dxk/ddx
         return y
\end{verbatim}
The iterated Aitken extrapolation applies Aitken extrapolation
repeatedly~\cite{Wen89}, as defined in the next function.

\begin{verbatim}
def repeatedAitken(x, exa, verbose=True):
    """
    Applies Aitken extrapolation repeatedly on the sequence x.
    If verbose, then the error with the exact value in exa is shown.
    Returns the last element.
    """
    cffs = x
    while len(cffs) > 2:
        a = Aitken(cffs)
        if verbose:
            print('on', len(cffs), ':', a[len(a)-1], end='')
            err = abs(a[len(a)-1] - exa)
            print(f' error :{err: .2e}')
        cffs = a
    return cffs[0]
\end{verbatim}
Executing {\tt repeatedAitken(ypt, 2.0)} produces the sequence
\begin{verbatim}
on 7 : (2.0137784911225802+0.004510752594407998j) error : 1.45e-02
on 5 : (2.0026132752890415+0.001430742263173204j) error : 2.98e-03
on 3 : (2.0008714460040284+0.000612101736300566j) error : 1.06e-03
\end{verbatim}
with similar results as {\tt repeatedAitken(xpt, 1.0)}.
The error (sum of errors on both $x$ and $y$ coordinates)
has decreased from {\tt 2.27e-01} to {\tt 1.69e-03}.
With relatively little effort, we gained two decimal places in the solution.

The rho algorithm and its iterated version produce similar results,
but its application is more delicate, as the default values of the
interpolation points need to take the values of the $t$ parameter
into account.

\section{Building and Installing phcpy}

Thanks to the efforts of Doug Torrance, both PHCpack and phcpy
can be installed with the package managers of Ubuntu.

Ad hoc makefiles, customized for Linux, Windows, and Mac OS X, 
were written during the development of PHCpack,
but became too cumbersome to maintain.
GPRbuild is the project manager of the GNAT toolchain,
which made the building of the executable {\tt phc}
and the library {\tt libPHCpack}, along with its components
in C and C++ (in particular: DEMiCs) more portable,
since version 2.4.85~\cite{Ver22}.
Since version 2.4.88, PHCpack can be built
via {\tt "alr get phcpack"}.

The development of phcpy started in python2, with the writing
of extension modules, which needed compilation,
and later adjustments for python3.
The compilation imposed restrictions for Windows,
as the Python interpreters on Windows are typically built with the
Microsoft compilers which are {\em not} interoperable with gcc.
Version 1.1.3 of phcpy (the third major rewrite of phcpy)
used the Ctypes module.  Once the {\tt libPHCpack} is in its proper place,
the instruction {\tt "pip install ."} in the folder where the {\tt setup.py}
is located will extend the Python interpreter with phcpy,
also on native Windows systems.

\section{Conclusions}

Extrapolating on Taylor series towards singular solutions,
this paper illustrates the application of numerical
analytic continuation to solving polynomial systems.
With {\tt phcpy}, the algorithms of~\cite{TVV20} and~\cite{VV22}
have become better accessible.

\bibliographystyle{plain}

\end{document}